# Scalar axion field of toroidal electromagnetic pulses

Wangke Yu[1], Nikitas Papasimakis[2], Nikolay I. Zheludev[1,2,3], Yijie Shen[1,4],

1. Centre for Disruptive Photonic Technologies & The Photonics Institute, School of Physical and Mathematical Sciences, Nanyang Technological University, Singapore 637371, Singapore
2. Optoelectronics Research Centre & Centre for Photonic Metamaterials, University of Southampton, Southampton SO17 1BJ, UK
3. Hagler Institute for Advanced Study, Texas A&M University, College Station, TX, USA
4. School of Electrical and Electronic Engineering, Nanyang Technological University, Singapore 639798, Singapore

**Abstract:**

Axion electrodynamics extends Maxwell's theory by postulating a hypothetical pseudoscalar axion field sourced by a scalar product of electric and magnetic fields. In this work, we demonstrate that a superposition of toroidal electromagnetic pulses propagating in free space naturally exhibits localized regions, where $\mathbf{E} \cdot \mathbf{B} \neq \mathbf{0}$. As a consequence of axion electrodynamics, these structured light pulses generate a space-time localized pseudoscalar field co-propagating with the pulses. This result should not be interpreted as a mechanism for generating axion particles by light, but rather as a consequence of adopting the axion electrodynamics extension to Maxwell's equations.

**Main:**

Axion electrodynamics, introduced by Frank Wilczek in 1987 [1], is an extension of Maxwell's theory, in which a hypothetical scalar axion field $a(\mathbf{r}, t)$ couples to the pseudoscalar electromagnetic invariant $\mathbf{E} \cdot \mathbf{B}$, and the modified Maxwell's equations are given by [1–4]:

$$\nabla \cdot \mathbf{E} = \rho - \kappa \nabla a \cdot \mathbf{B}, \tag{1}$$

$$\nabla \times \mathbf{E} = -\partial_t \mathbf{B}, \tag{2}$$

$$\nabla \cdot \mathbf{B} = 0, \tag{3}$$

$$\nabla \times \mathbf{B} - \partial_t \mathbf{E} = \mathbf{J} + \kappa (\dot{a} \mathbf{B} + \nabla a \times \mathbf{E}), \tag{4}$$

where $\rho$ and $\mathbf{J}$ denote conventional charge and current densities, $t$ is time, and $\kappa$ is the axion-photon coupling parameter. When $a(r,t)$ is uniform and static, Eqs. (1) and (4) reduce to the conventional Maxwell equations. In free space, the axion field motion obeys a Klein–Gordon equation, where $\mathbf{E} \cdot \mathbf{B}$ is the local source term [1–4],

$$\ddot{a} - \nabla^2 a + m_a^2 a = -\kappa\, \mathbf{E} \cdot \mathbf{B}. \tag{5}$$

In the absence of the $\mathbf{E} \cdot \mathbf{B}$ drive, Eq. (5) reduces to free massive field dynamics, and Eqs. (1)-(4) reduce to the conventional Maxwell equations. This coupling underpins the axion–photon interaction [2-5] and motivates experimental searches [6,7] for axion particles. The values of axion particle mass $m_a$ and coupling strength $\kappa$ remain unknown, the theoretical estimates of which and experimental search for axion dark matter are still emerging topics [7-10]. Closely related effective axion-like electrodynamics has also arisen in the description of topological magnetoelectric media and quantum materials [11-15], and, recently, photonic materials with engineered axion response [16, 17].

The space-time localized $\mathbf{E} \cdot \mathbf{B} \neq 0$ excitations in the free space are of particular interest because they supply compact, propagating source terms for the axion-electrodynamics field equation. In a conventional plane electromagnetic wave, the field is always of $\mathbf{E} \perp \mathbf{B}$ ($\mathbf{E} \cdot \mathbf{B} = 0$), while, nonzero $\mathbf{E} \cdot \mathbf{B}$ can be produced by the interference of several plane waves of different propagating directions or tight focusing of light [18, 19]. Spatial localized nonzero $\mathbf{E} \cdot \mathbf{B}$ structures can also be found in surface evanescent field and their importance for axion electrodynamics was recently discussed [20]. However, how a space-time localized nonzero $\mathbf{E} \cdot \mathbf{B}$ can propagate in a wave packet is still unknown.

In 1996, Hellwarth and Nouchi theoretically identified an exact non-transverse space-time non-separable type of electromagnetic pulse with toroidal topology [21]. These pulses, often called focused or flying doughnuts (FDs), have since been realized experimentally in the optical, terahertz and microwave-domain [22-25]. They are few-cycle transverse-electric (TE) or transverse-magnetic (TM) pulses with nonseparable spatiotemporal and polarization structure, and they can support nontrivial physical features such as skyrmionic field textures, fractal-like singularities, isodiffraction and energy backflows [24-28].

In this work, we show that the interference of toroidal pulses provides a platform for generating a propagating $\mathbf{E} \cdot \mathbf{B}$ source in free space. While a pure TE or TM toroidal pulse does not exhibit regions with nonzero $\mathbf{E} \cdot \mathbf{B}$, the interference of TE and TM pulses generates such zones and thereby drives the scalar sector of axion electrodynamics in a controlled and analytically transparent way, as illustrated in Fig.1.

Figure 1(a) shows the non-transverse electric field of a propagating TM toroidal pulse at different propagation distances ($z = -z_0, 0, +z_0$, where $z_0$ is the Rayleigh range) [21]. The field comprises coupled radial $E_r$ and longitudinal $E_z$ components, and the vector electric field distribution in the $(x, z)$ plane at focus ($t = 0$) is shown in Fig. 1(b). The grey quiver plot indicates the local electric-field direction, while the thin green contours mark iso-surfaces of the electric field modulus. Together they reveal the circulation of the field along the poloidal direction, highlighted schematically by the bold green loops. This poloidal electric field loop is shown explicitly in the field line rendering of the TM family in Fig. 1(c). In the TE family, the electric and magnetic sectors are exchanged relative to the TM family, see Fig. 1(d). Fig. 1(e) presents a schematic representation of the superposition of TM and TE modes, where the mixed TE/TM terms generate a localized region with nonzero $\mathbf{E} \cdot \mathbf{B}$.

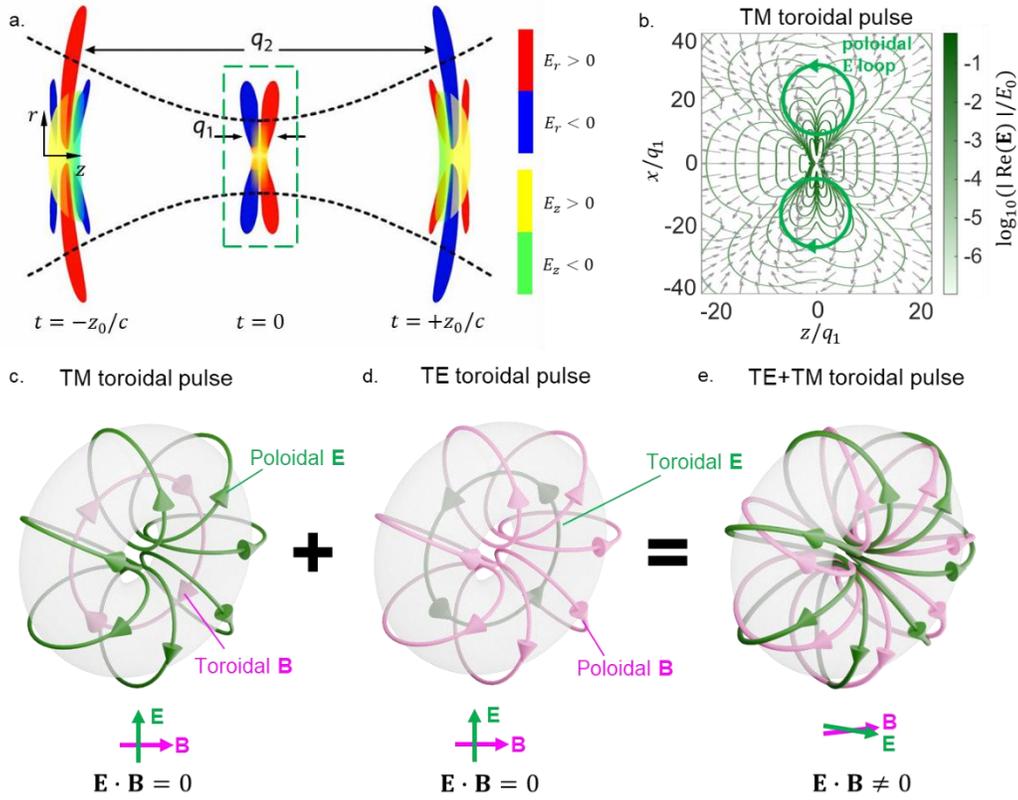

**Figure 1. Interference of toroidal pulses - a source of localized $\mathbf{E} \cdot \mathbf{B}$. (a)** The radial $E_r$ (red/blue) and longitudinal $E_z$ (yellow/green) electric-field components of TM toroidal pulse at three axial positions (e.g., $z = -z_0, 0, +z_0$), plotted at a fixed isovalue (colors indicate sign only). The dashed curves indicate the pulse's envelope, and $q_1$ and $q_2 = 2z_0$ define the characteristic wavelength and focal-depth scales (see Eqs. (6)-(8)). **(b)** The $(x, z)$ section of the same pulse ($t = 0$): arrows show the direction of $\text{Re}(\mathbf{E})$, and the colormap shows $\log_{10}(|\text{Re}(\mathbf{E})|/E_0)$, where $E_0$ is the peak field amplitude. The green curves correspond to the green field-line of $\mathbf{E}$-field in panel (c). **(c)** Field-lines of a TM toroidal pulse. **(d)** Field-lines of a TE toroidal pulse. Relative to the TM case in panel (c), the electric and magnetic fields are interchanged, while the azimuthal component reverses sign. **(e)** Interference of TE and TM toroidal pulses. The individual TE and TM toroidal pulses do not exhibit $\mathbf{E} \cdot \mathbf{B} \neq 0$ on their own, but their hybridization produces a localized region of $\mathbf{E} \cdot \mathbf{B} \neq 0$.

We evaluate the free-space invariant $\mathbf{E} \cdot \mathbf{B}$ of the superposed TE and TM toroidal pulses and use it as the driving term in Eq. (5). We do this in the weak-coupling, negligible-backreaction approximation, assuming $\rho = 0, \mathbf{J} = 0$, and neglecting the axion-induced terms on the right-hand sides of Eqs. (1) and (4). Under these assumptions, Eqs. (1)–(4) reduce to the ordinary Maxwell equations for a prescribed toroidal electromagnetic background, while Eq. (5) is solved for the scalar field $a$ driven by the $\mathbf{E} \cdot \mathbf{B}$ source. To describe the $\mathbf{E} \cdot \mathbf{B}$ source we use the analytic representation of the TM and TE toroidal pulses obtained from the scalar generating function in the $(r,z)$ cylindrical coordinate system [21,22,24]:

$$E_z^{TM} = cB_z^{TE} = -4f_0\sqrt{(\mu_0/\varepsilon_0)}[r^2 - (q_1 + i\tau)(q_2 - i\sigma)]/[r^2 + (q_1 + i\tau)(q_2 - i\sigma)]^3, \quad (6)$$

$$E_r^{TM} = c\,B_r^{TE} = 4if_0\sqrt{(\mu_0/\varepsilon_0)}r(q_2 - q_1 - 2iz)/[r^2 + (q_1 + i\tau)(q_2 - i\sigma)]^3, \quad (7)$$

$$B_\theta^{TM} = -\frac{1}{c}E_\theta^{TE} = 4i\mu_0 f_0 r(q_1 + q_2 - 2ict)/[r^2 + (q_1 + i\tau)(q_2 - i\tau)]^3. \quad (8)$$

where, $\tau = z - ct$ and $\sigma = z + ct$; $\mu_0$ and $\varepsilon_0$ are the vacuum permeability and permittivity, respectively; $t$ is time; $f_0$ is a normalization constant; and $q_1$ and $q_2$ set the effective wavelength and Rayleigh range ($z_0 = q_2/2$) of the pulse, respectively [25,27,31]. The complex fields in Eqs. (6)–(8) are analytic representations of the toroidal pulses. Their real and imaginary parts represent distinct physical solutions of Maxwell's equations, corresponding to toroidal pulses of single-cycle and $1\frac{1}{2}$-cycle duration at focus, respectively. In this work, we consider the solution obtained from the real part; since both solutions share the same polarization structure, analogous results are expected for the solution obtained from the imaginary part. The electric and magnetic fields of the solution, derived from the real part, are denoted by $\mathrm{Re}(\mathbf{E})$ and $\mathrm{Re}(\mathbf{B})$, respectively. And the associated second Lorentz invariant is given by $\mathrm{Re}(\mathbf{E}) \cdot \mathrm{Re}(\mathbf{B})$.

The superposition of TE and TM toroidal pulses produces a self-dual electromagnetic configuration. The resulting field is invariant under the duality transformation and exhibits nonzero electric and magnetic projections along the propagation direction. The resulting localized, propagating source term $\mathbf{E} \cdot \mathbf{B}$ source term has a sign and symmetry that can be tuned by the relative phase $\delta$. Such superpositions of toroidal pulses have been generated experimentally in the optical [29, 30] and microwave parts of the spectrum [31].

Figure 2 illustrates the scalar-field solutions $a(z,x;t)$ obtained from Eq. (5) for two relative phases of the TE/TM hybrid. The figure shows that the driven scalar-field packet remains localized and co-propagates with the TE/TM hybrid.

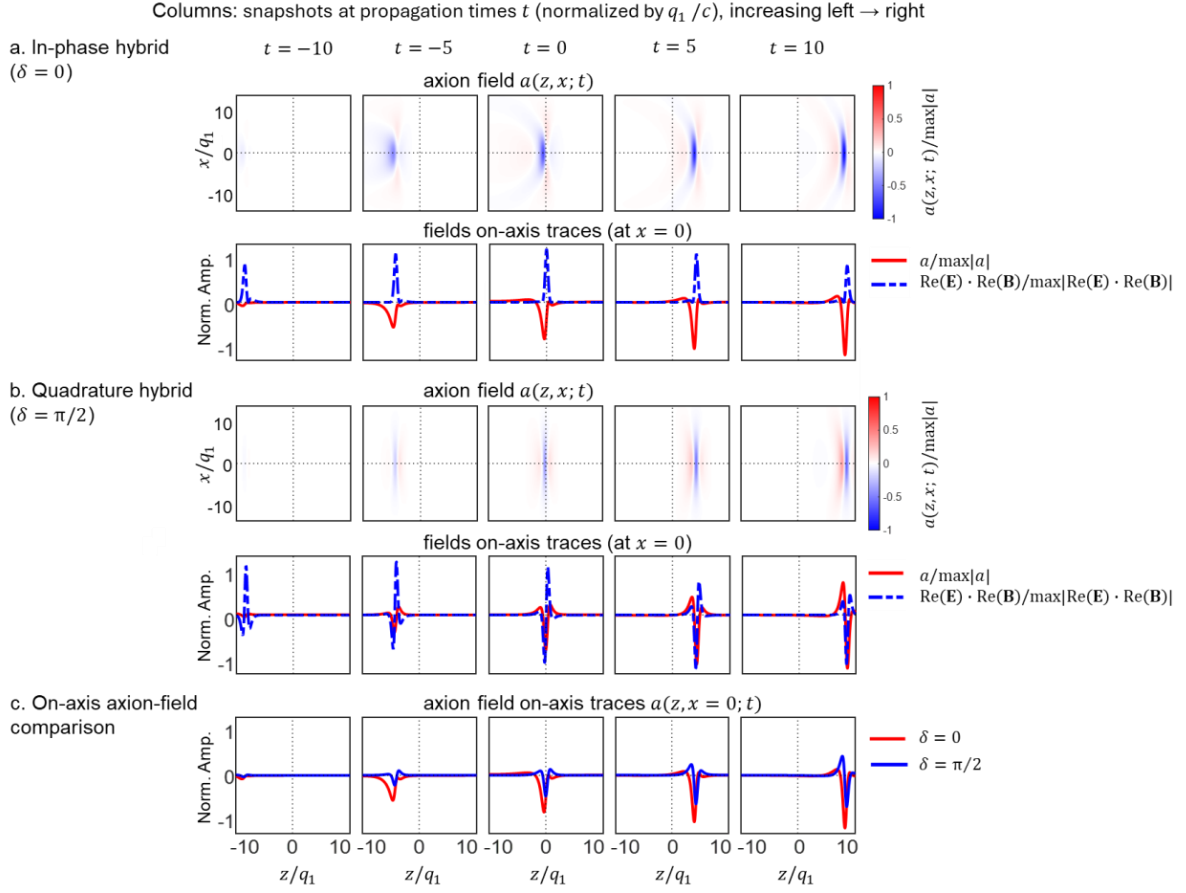

**Figure 2.** Scalar axion field driven by hybridized TE and TM toroidal pulses. **(a)** The scalar field $a$ obtained from Eq. (5) is shown for a superposition of TM and TE pulses interfering without a phase delay, $\delta = 0$. **(b)** The scalar field $a$ obtained from Eq. (5) is shown for a superposition of TM and TE pulses interfering with a phase delay, $\delta = \pi/2$. The scalar field profile $a(z, x; t)$ normalized by $\max |a|$. **(c)** On-axis comparison of $a(z, x = 0, t)$ for $\delta = 0$ and $\delta = \pi/2$.

In conclusion, we have demonstrated that the superposition of TE and TM Hellwarth-Nouchi toroidal pulses results in a propagating localized pseudoscalar $\mathbf{E} \cdot \mathbf{B}$ excitation that, within the axion extension of Maxwell electrodynamics, drives a classical pseudoscalar axion field co-propagating with the pulses. Our results on generating a spacetime-localized pseudoscalar field should not be interpreted as a mechanism for photon–axion scattering, but rather as a consequence of adopting the axion electrodynamics extension to Maxwell's equations.